\documentclass[conference]{IEEEtran}

\usepackage{amssymb}
\usepackage[cmex10]{amsmath}
\usepackage{stfloats}
\usepackage{graphicx}
\usepackage{subfigure}
\usepackage{tabularx}
\usepackage{epsfig,epsf,color,balance,cite}
\usepackage{verbatim}

\newtheorem{theorem}{Theorem}
\newtheorem{lemma}{Lemma}
\usepackage{algorithm}
\usepackage{algpseudocode}
\usepackage{amsmath}

\begin{document}

\title{Energy Efficient Resource Allocation for Mobile-Edge Computation Networks with NOMA}

\author{
\IEEEauthorblockN{Zhaohui Yang,
Jiancao Hou
                  and Mohammad Shikh-Bahaei
                  }
                  
\IEEEauthorblockA{Centre for Telecommunications Research, Department of Informatics, King's College London, U.K.}
\IEEEauthorblockA{E-mail: \{yang.zhaohui, jiancao.hou, m.sbahaei\}@kcl.ac.uk} 
}

\maketitle

\begin{abstract}

This paper investigates an uplink non-orthogonal multiple access (NOMA)-based mobile-edge computing (MEC) network. Our objective is to minimize the total energy consumption of all users including transmission energy and local computation energy subject to computation latency and cloud computation capacity constraints. We first prove that the total energy minimization problem is a convex problem, and it is optimal to transmit with maximal time. Then, we accordingly proposed an iterative algorithm with low complexity, where closed-form solutions are obtained in each step. The proposed algorithm is successfully shown to be globally optimal. Numerical results show that the proposed algorithm achieves better performance than the conventional methods.
\end{abstract}

\IEEEpeerreviewmaketitle

\section{Introduction}
With the rapid development of intelligent communications \cite{vaezi2018multiple,nehra2010cross,xu2017corticomuscular,8436454,7106478,towhidlou2018improved},
mobile-edge computing (MEC) has been deemed as a promising technology for future communications due to that it can improve the computation capacity of users in applications, such as, augmented reality (AR) \cite{8016573}.
With MEC, users can offload the tasks to the MEC servers that are located at the edge of the network.
Since the MEC servers can be deployed near to the users, network with MEC can provide users with low latency and low energy consumption \cite{7906521,ming2017,raman2017wi,8422519}.

The basic idea of MEC is to utilize the powerful computing facilities within the radio access network, such as the MEC server integrated into the base station (BS).
Users can benefit from offloading the computationally intensive tasks to the MEC server.
There are two operation modes for MEC, i.e., partial and binary computation offloading.
In partial computation offloading, the computation tasks can be divided into two parts, where one part is locally executed and the other part is offloaded to the MEC server \cite{8006982,8254208,7762913,7929399,8240666,8168252,7842016}.
In binary computation offloading, the computation tasks are either locally executed or offloaded to the MEC server \cite{6574874}.

Recently, non-orthogonal multiple access (NOMA) has been recognized as a potentional technology for the next generation mobile communication networks to tackle the explosive growth of data traffic \cite{saito2013non,Yang2018Power,Zhiguo2017Survey,Yang2017Fair,7263349}.
Due to superposition coding at the transmitter and successive interference cancelation (SIC) at the receiver, NOMA can achieve higher spectral efficiency than conventional orthogonal multiple access (OMA), such as time division multiple access (TDMA) and orthogonal frequency division multiple access (OFDMA).
Many previous contributions \cite{7906521,8006982,8254208,Yang2017Energy,7762913,7929399,8240666,8168252,7842016,Yang2018EEIoT} only considered OMA.
Motivated by the benefits of NOMA over OMA, a NOMA-based MEC network was investigated in \cite{8269088}, where users simultaneously offload their computation tasks to the BS and the BS uses SIC for information decoding.
Besides, both NOMA uplink and downlink transmissions were applied to MEC \cite{zhi2018ImNOMAMEC}, where analytical results were developed to show that  the latency and energy consumption can be reduced by applying NOMA-based MEC offloading.
Time and energy minimization were respectively optimized in \cite{Ding2018Delay} and \cite{Ding2018Joint} for NOMA-based MEC networks with different computation deadline requirements for different users.
However, \cite{8269088,zhi2018ImNOMAMEC,Ding2018Delay,Ding2018Joint} only considered one group of users forming NOMA and ignored the time allocation among different groups of users forming NOMA.
Since each resource is recommended to be multiplexed by small number of users (for example, two users) due to decoding complexity and error propagation \cite{6464495}, it is of importance to investigate the  resource allocation among different groups of users forming NOMA.

In this paper, we investigate the resource allocation for an uplink NOMA-based MEC network.
The main contributions of this paper are summarized as follows:
\begin{enumerate}
  \item  The total energy consumption of all users is formulated for an uplink NOMA-based MEC network via optimizing transmission power, offloading data and time allocation.
      Different from \cite{8269088} and \cite{zhi2018ImNOMAMEC}, time allocation for different groups is investigated in this paper, where two users are paired in each group to perform NOMA.
  \item  The total energy minimization problem is proved to be a convex one.
  Besides, it is also shown that transmitting with maximal time is optimal in energy saving.
\item Based on the optimal conditions, an iterative algorithm is accordingly proposed, where closed-form expressions are obtained in each step for optimizing time allocation or offloading data.
The proposed iterative algorithm with low complexity is successfully proved to be globally optimal.
\end{enumerate}

The rest of the paper is organized as follows.
In Section $\text{\uppercase\expandafter{\romannumeral2}}$, we introduce the system model and formulate the total energy minimization problem.
Section $\text{\uppercase\expandafter{\romannumeral 3}}$ provides the optimal conditions and an iterative algorithm.
Some numerical results are shown in Section $\text{\uppercase\expandafter{\romannumeral 4}}$
and conclusions are finally drawn in Section $\text{\uppercase\expandafter{\romannumeral5}}$.

\section{System Model and Problem Formulation}

\begin{figure}
\centering
\includegraphics[width=3.35in]{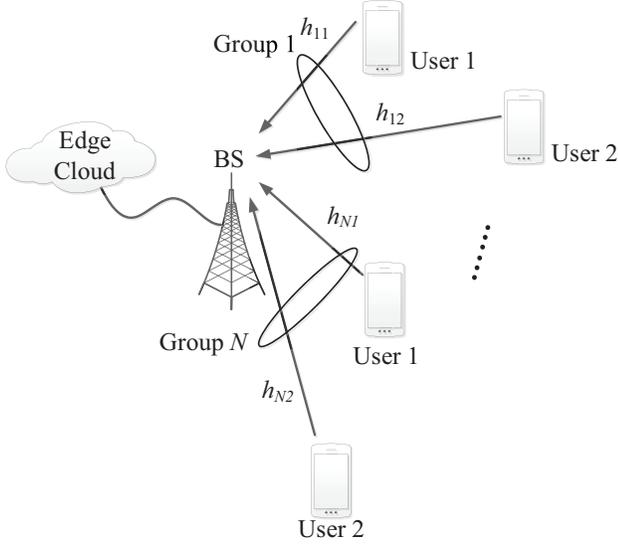}
\vspace{-0.5em}
\caption{Multi-user MEC network.}\label{sys1fig1}
\vspace{-0.5em}
\end{figure}

\begin{figure}
\centering
\includegraphics[width=3.35in]{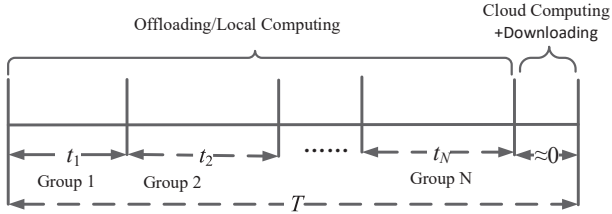}
\vspace{-0.5em}
\caption{Transmission period.}\label{sys1fig2}
\vspace{-0.5em}
\end{figure}

Consider a NOMA-enabled MEC network with $2N$ users and one BS that is the gateway of an edge cloud, as shown in Fig.~\ref{sys1fig1}.
All $2N$ users are classified into $N$ groups with two users in each group.
Let $\mathcal N=\{1, 2, \cdots, N\}$ denote the set of all groups.
In each group, these two users simultaneously transmit data to the BS at the same frequency by using NOMA.
We consider TDMA scheme for users in different group, as shown in Fig.~\ref{sys1fig2}.

The BS schedules the users to completely or partially offload tasks.
The users with complete or partial offloading respectively offload a fraction of or all input data to the BS, while the users with partial or no offloading respectively compute a fraction of or all input data using local central processing unit (CPU).
The channel is assumed to be frequently flat.
Due to the small latency of cloud computing and small sizes of computation results, the time of cloud computing and downloading from the BS is negligible compared to the time of mobile offloading and local computing \cite{7842016}.

The BS is assumed to have the perfect information of the channels, local computation capabilities and input data sizes of all users.
Using this information, the BS determines the transmission power, the offloaded data, and the fraction of offloading time.

\subsection{Task Computing Model}
The local computing model is described as follows.
Since only $d_{ij}$ bits are offloaded to the BS, the remaining $R_{ij}-d_{ij}$ bits are needed to be computed locally at user $j$ in group $i$.
Based on the local computing model in  \cite{7842016}, the total energy consumption for local computation at user $j$ in group $i$ is given by
\begin{equation}\label{sys1eq1}
E_{ij}^{\text{Loc}} = (R_{ij}-d_{ij}) C_{ij}P_{ij}, \quad \forall i\in\mathcal N, j=1,2,
\end{equation}
where $C_{ij}$ is the number of CPU cycles required for computing 1-bit input data at user $j$ in group $i$,
and $P_{ij}$ stands for the energy consumption per cycle for local computing at this user.

Let $F_{ij}$ denote the computation capacity of user $j$ in group $i$, which is measured by the number of CPU cycles per second.
Denoting $T$ as the maximal latency of all users, we can obtain the following local computation latency constraints
\begin{equation}\label{sys1eq1_2}
(R_{ij}-d_{ij}) C_{ij}\leq F_{ij} T, \quad \forall i\in\mathcal N, j=1,2,
\end{equation}
which can be equivalent to
\begin{equation}\label{sys1eq1_22}
d_{ij} \geq \frac{R_{ij} C_{ij}-F_{ij} T}
{C_{ij}},
 \quad \forall i\in\mathcal N, j=1,2.
\end{equation}

It is also assumed that the edge cloud has finite computation capacity $F$.
As a result, the offloading data of all users should satisfy the following computation constraint:
\begin{equation}\label{sys1eq1_3}
\sum_{i=1}^N \sum_{j=1}^2 d_{ij} C_{ij}\leq F.
\end{equation}

\subsection{Offloading Model}
Denotes the bandwidth of the network by $B$, and the power spectral density of the additive white Gaussian noise by $\sigma^2$.
Let $h_{ij}$ denote the channel gain between user $j$ in group $i$ and the BS.
Without loss of generality, the uplink channels between users in group $i$ and the BS are sorted as $h_{i1} \geq  h_{i2}$, $\forall i\in \mathcal N$.

Users in each group will be assigned with a fraction of time to use the whole bandwidth.
The time allocated with users in group $i$ is denoted by $t_i$.
To meet the uploaded data demand, we have
\begin{equation}\label{sys1eq2_1}
d_{i1}=r_{i1}t_i, d_{i2}=r_{i2}t_i, \quad \forall i \in \mathcal N,
\end{equation}
where
\begin{equation}\label{sys1eq2}
r_{i1}=B\log_2 \left( 1+\frac {p_{i1} h_{i1} } {\sigma^2B +p_{i2} h_{i2} } \right),
\end{equation}
and
\begin{equation}\label{sys1eq3}
r_{i2} =B \log_2 \left( 1+\frac {p_{i2} h_{i2} } {\sigma^2B  } \right).
\end{equation}
Note that the BS detects the messages of two users via NOMA technique, i.e., the BS first detects the message of strong user 1 and then detects the message of weak user 2 with SIC \cite{7022998,8294259,8320533}.
As a result, the achievable rates of user 1 and 2 in group $i$ can be given by (\ref{sys1eq2}) and (\ref{sys1eq3}), respectively.
Substituting $r_{i1}=d_{i1}/t_i$ and
$r_{i2}=d_{i2}/t_i$ obtained from (\ref{sys1eq2_1}) into (\ref{sys1eq2}) and (\ref{sys1eq3}) yields
\begin{equation}
 p_{i1}=a_{i1}B\left( 2^{\frac{d_{i1}+d_{i2}}{Bt_i}}-2^{\frac{d_{i2}}{Bt_i}}\right),   p_{i2}=a_{i2}B \left( 2^{\frac{d_{i2}}{Bt_i}}-1\right),\!\!  \label{sys1eq5}
\end{equation}
where
\begin{equation}\label{sys1eq6_1}
a_{i1}=\frac{\sigma^2}
{h_{i1}},a_{i2}=\frac{\sigma^2}{h_{i2}}.
\end{equation}
Based on (\ref{sys1eq5}), the energy consumption for offloading at users in group $i$ is given by
\begin{eqnarray}\label{sys1eq6}
\!\!\!\!\!\!\!\!E_{i}^{\text{Off}}&&\!\!\!\!\!\!\!\!\!\!\!=\sum_{j=1}^2 p_{ij}t_{i}\nonumber\\
&&\!\!\!\!\!\!\!\!\!\!\!
=Bt_i\!\left(\!a_{i1} 2^{\frac{d_{i1}+d_{i2}}{Bt_i}}
+(a_{i2}-a_{i1}) 2^{\frac{d_{i2}}{Bt_i}}
-a_{i2}\!\right)\!.
\end{eqnarray}

\subsection{Problem Formulation}
Now, it is ready to formulate the sum user energy minimization problem as:
\begin{subequations}\label{sys1min1}
\begin{align}
\mathop{\min}_{\pmb d, \pmb t}\;
 \quad&  \sum_{i=1}^N Bt_i\!\left(\!a_{i1} 2^{\frac{d_{i1}+d_{i2}}{Bt_i}}
+(a_{i2}-a_{i1}) 2^{\frac{d_{i2}}{Bt_i}}
-a_{i2}\!\right)
 \nonumber\\&
 +\sum_{i=1}^N \sum_{j=1}^2  (R_{ij}-d_{ij}) C_{ij}P_{ij}
  \\
\textrm{s.t.}\quad\qquad \!\!\!\!\!\!\!\!\!
&\sum_{i=1}^N t_i \leq T \\
& \sum_{i=1}^N \sum_{j=1}^2 d_{ij} C_{ij}\leq F\\
&  D_{ij} \leq d_{ij} \leq R_{ij},t_i\geq 0, \quad \forall i\in\mathcal N, j=1,2,
\end{align}
\end{subequations}
where $\pmb d=[d_{11},d_{12},\cdots,d_{N1},d_{N2}]$,  $\pmb t=[t_1, \cdots,t_N]$ and $D_{ij}=\max\{\frac{R_{ij} C_{ij}-F_{ij} T}
{C_{ij}},0\}$.
The objective function (\ref{sys1min1}a) represents the total energy consumption of all users including both offloading energy and computing energy.
The time division constraint is shown in (\ref{sys1min1}b).
Constraint (\ref{sys1min1}c) shows the maximal computation capacity limit.
Constraints (\ref{sys1min1}d) ensure that the local computation can be finished in time constraint $T$ for all users.

\section{Optimal Solution}

In this section, we first provide the optimal conditions of sum energy minimization problem (\ref{sys1min1}), and then accordingly propose an iterative algorithm to obtain the optimal solution of problem (\ref{sys1min1}).

\subsection{Optimal Conditions}
Before solving problem (\ref{sys1min1}), several characteristics are provided as follows.
\begin{lemma}
Problem (\ref{sys1min1}) is a convex problem.
\end{lemma}

\itshape \textbf{Proof:}  \upshape
Please refer to Appendix A.
\hfill $\Box$

\begin{lemma}
It is optimal to transmit with the maximal time, i.e., $\sum_{i=1}^N t_i^*=T$ for problem (\ref{sys1min1}).
\end{lemma}

\itshape \textbf{Proof:}  \upshape
Please refer to Appendix B.
\hfill $\Box$

Lemma 1 shows that problem (\ref{sys1min1}) is a convex problem, which can be effectively solved to its optimality.
According to Lemma 2, transmitting with maximal time is always energy efficient.
The reason is that, as the transmission time increases, the required power decreases and then the product of time and power, which can be viewed as the consumed energy, also decreases.

\subsection{Iterative Algorithm}
Even problem (\ref{sys1min1}) is convex, it is difficult to obtain the optimal solution of problem (\ref{sys1min1}) in closed form due to the fact that the objective function (\ref{sys1min1}a) couples both offloading data $\pmb d$ and time allocation $\pmb t$.
In the following, we propose an iterative algorithm via optimizing time allocation $\pmb t$ with fixed offloading data $\pmb d$ and solving offloading data $\pmb d$ with fixed time allocation $\pmb t$, where the closed-form solution can be fortunately obtained in each step.

\begin{theorem}
With fixed offloading data $\pmb d$, the optimal time allocation of problem (\ref{sys1min1}) is
\begin{equation}\label{IA3eq1}
t_i^*=[g_i'^{-1}(-\alpha)]^+,\quad \forall i \in \mathcal N,
\end{equation}
where  $g_i'^{-1}(x)$ is the inverse function of $g_i'(x)$ defined in (\ref{AppenBeq2}), $[x]^+=\max\{x,0\}$, and $\alpha$ satisfies
\begin{equation}\label{IA3eq2}
\sum_{i=1}^N [g_i'^{-1}(-\alpha)]^+=T.
\end{equation}
\end{theorem}
\itshape \textbf{Proof:}  \upshape
Please refer to Appendix C.
\hfill $\Box$

Before presenting Theorem 2 about the optimal offloading data, we define  \begin{equation}\label{IA3eq3}
 d_{i1}(\beta)\!=\!\left.Bt_i\log_2 \!\left(\!
\frac{ (a_{i2}\!-\!a_{i1})(  P_{i1}C_{i1}\!-\!\beta C_{i1})}
{a_{i1}(\beta(C_{i1}\!-\!C_{i2})\!-\!P_{i1}C_{i1}\!+\!P_{i2}C_{i2} )}
\!\right) \!\right|_{D_{i1}}^{R_{i1}}\!,
\end{equation}
and
\begin{equation}\label{IA3eq5}
d_{i2}(\beta)=\left.Bt_i\log_2 \!\left(\!
\frac{\beta(C_{i1}\!-\!C_{i2})\!-\!P_{i1}C_{i1}\!+\!P_{i2}C_{i2}}
{(\ln 2)(a_{i2}\!-\!a_{i1})}
\!\right) \right|_{D_{i2}}^{R_{i2}}\!.\!
\end{equation}
where $a|_b^c=\min\{\max\{a,b\},c\}$.

\begin{theorem}
1) If $\sum_{i=1}^N \sum_{j=1}^2 d_{ij}(0)C_{ij}\leq F$,
 the optimal offloading data of problem (\ref{sys1min1}) with fixed time allocation $\pmb t$, is given by
 \begin{equation}\label{IA3eq6}
 d_{i1}^*=d_{i1}(0),d_{i2}^* =d_{i2}(0), \quad \forall i\in \mathcal N.
\end{equation}
2) If $\sum_{i=1}^N \sum_{j=1}^2 d_{ij}(0)C_{ij} > F$,
 the optimal offloading data of problem (\ref{sys1min1}) with fixed time allocation $\pmb t$, is
\begin{equation}\label{IA3eq6_2}
 d_{i1}^*=d_{i1}(\beta),d_{i2}^* =d_{i2}(\beta), \quad \forall i\in \mathcal N,
\end{equation}
 where $\beta$ satisfies
 \begin{equation}\label{IA3eq7}
 \sum_{i=1}^N \sum_{j=1}^2 d_{ij}(\beta) C_{ij}=F.
 \end{equation}
\end{theorem}
\itshape \textbf{Proof:}  \upshape
Please refer to Appendix D.
\hfill $\Box$

By iteratively solving time allocation problem and  offloading data problem, the algorithm that solves problem (\ref{sys1min1}) is given in Algorithm~1.

\begin{algorithm}[h]
\caption{: Iterative Time and Offloading Data Allocation }
\begin{algorithmic}[1]
\State Initialize ${\pmb d}^{(0)}=[D_{11}, D_{12} \cdots, D_{N1}, D_{N2}]$, ${\pmb t}^{(0)}=[T/N, \cdots,$ $ T/N]$,  the tolerance $\xi$, the iteration number $l=0$ and the maximal iteration number $L_{\max}$.
\State Compute the objective value $V_{\text {obj}}^{(0)}=\bar V({\pmb d}^{(0)},  {\pmb t}^{(0)})$, where $\bar V({\pmb d},  {\pmb t})$ equals to the objective function (\ref{sys1min1}a).
\State With given offloading data $ {\pmb d}^{(l)}$, obtain the optimal time allocation $\pmb t^{(l+1)}$ according to Theorem 1.
\State With given time allocation $ {\pmb t}^{(l+1)}$, obtain the optimal offloading data  $\pmb d^{(l+1)}$ according to Theorem 2.
\State Compute the objective value $V_{\text {obj}}^{(l+1)}=\bar V({\pmb d}^{(l+1)}, {\pmb t}^{(l+1)})$.
If $|V_{\text {obj}}^{(t+1)}-V_{\text {obj}}^{(t)}|/V_{\text {obj}}^{(t)}<\xi$ or $l>L_{\max}$, terminate.
Otherwise, set $l=l+1$ and go to step 3.
\end{algorithmic}
\end{algorithm}

\subsection{Optimality and Complexity Analysis}

\begin{theorem}
The proposed Algorithm 1 always converges to the global optimum of problem (\ref{sys1min1}).
\end{theorem}
\itshape \textbf{Proof:}  \upshape
Please refer to Appendix E.
\hfill $\Box$

Note that the proposed iterative Algorithm 1 yields the globally optimal solution to convex problem (\ref{sys1min1}) thanks to the fact that constraints (\ref{sys1min1}b)-(\ref{sys1min1}d) are not coupled with offloading data $\pmb d$ and time $\pmb t$, i.e., constraint (\ref{sys1min1}b) only involves time $\pmb t$, while only offloading data $\pmb d$ appears in constraint (\ref{sys1min1}c) and constraints (\ref{sys1min1}d) are box constraints.
As a result, the proposed iterative Algorithm 1 always converges to a local optimal solution, i.e., the globally optimal solution to the original convex problem (\ref{sys1min1}).

According to Algorithm 1, the major complexity lies in solving the offloading data allocation of problem (\ref{sys1min1}) with fixed time allocation.
From Theorem 2, the main complexity of obtaining the optimal offloading data lies in solving equation (\ref{IA3eq7}) by using the one-dimension search method with complexity $\mathcal O(NK)$, where $K$ denotes the number of iterations for the one-dimension search method.
As a result, the total complexity of the proposed Algorithm 1 is $\mathcal O(L_{\text{it}}NK)$, where $L_{\text{it}}$ is the number of iterations for iteratively optimizing time allocation and offloading data.
Due to the fact that the dimension of the variables in problem (\ref{sys1min1}) is $3N$, the complexity of solving
problem (\ref{sys1min1}) by using the standard interior point method is
$\mathcal O(L_{\text{ip}}N^3)$ \cite[Pages~487, 569]{boyd2004convex},
where $L_{\text{ip}}$ denotes the number of iterations for the interior point method.

\section{Numerical Results}

In this section, numerical results are presented  to evaluate the performance of the proposed algorithm.
The NOMA-enabled MEC network consists of $2N=30$ users.
The path loss model is $128.1+37.6\log_{10} d$ ($d$ is in km)
and the standard deviation of shadow fading is $4$ dB \cite{access2010further}.
In addition, the bandwidth of the network is $B=10$ MHz, and the noise power density is  $\sigma^2=-169$ dBm/Hz.
For MEC parameters, the data size and the required number of CPU cycles per bit are  set to follow equal distributions with $R_{ij} \in[100,500]$ Kbits and $C_{ij}\in[500, 1500]$ cycles/bit.
The CPU computation of each user is set as the same $F_{ij}=1$ GHz and the local computation energy per cycle for each user is also set as equal $P_{ij}=10^{-10}$ J/cycle for all $i \in \mathcal N$ and $j=1,2$.
Unless specified otherwise, the system parameters are set as time solt duration $T=0.1 $ s, and the edge computation capacity $F=6\times 10^9$ cycles per slot.

\begin{figure}
\centering
\includegraphics[width=3.0in]{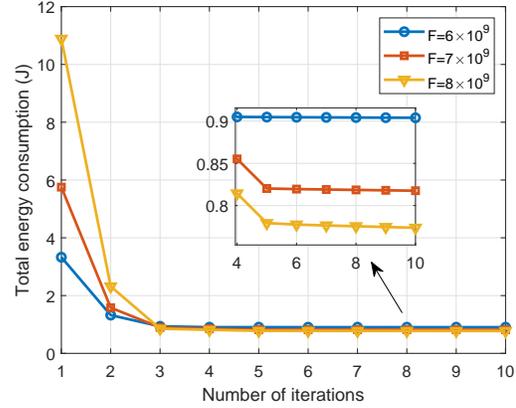}
\vspace{-0.5em}
\caption{Convergence behaviour of the proposed algorithm under different cloud computation capacities.}\label{fig1}
\vspace{-0.5em}
\end{figure}

Fig. \ref{fig1} illustrates the convergence behaviours for the proposed algorithm under different cloud computation capacities.
It can be seen that the proposed algorithm converges rapidly, and only three times are sufficient to converge, which shows the effectiveness of the proposed algorithm.

\begin{figure}
\centering
\includegraphics[width=3.0in]{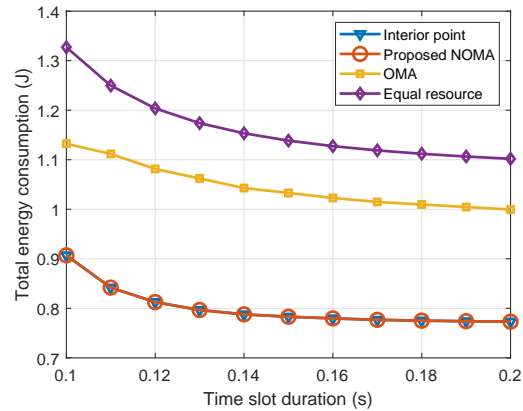}
\vspace{-0.5em}
\caption{Total energy consumption versus time slot duration.}\label{fig2}
\vspace{-0.5em}
\end{figure}

We compare the total energy consumption performance of the proposed algorithm (labelled as `Proposed NOMA') with the interior point method to solve convex problem (\ref{sys1min1}) by using matlab toolbox, the conventional optimal algorithm for OMA-based MEC networks \cite{7842016} (labelled as `OMA'), and the equal resource allocation algorithm where equal time duration is allocated for different groups and the offloading data is optimally allocated (labelled as `Equal resource').

The total energy consumption versus time slot duration is depicted in Fig.~\ref{fig2}.
From this figure, we find that the total energy consumption decreases with time slot duration.
This is due to the fact that transmitting with long time is energy efficient according to Lemma 2.
It can be shown that the proposed algorithm yields almost the same performance as the interior point method.
This is because the proposed total energy minimization problem (\ref{sys1min1}) is a convex problem, both the proposed algorithm and the interior point method can obtain the same globally optimal solution, which verifies the theoretical analysis in Theorem 3.
It is also found the proposed algorithm yields better performance than the OMA and equal resource schemes.
Compared with OMA, NOMA reduces the total energy consumption of all users at the cost of adding computing complexity at the BS due to SIC.
Since only simple equal time allocation is assumed in equal resource scheme, the proposed algorithm jointly optimizes both time allocation and offloading data, which results in lower energy consumption in the proposed algorithm.

\begin{figure}
\centering
\includegraphics[width=3.0in]{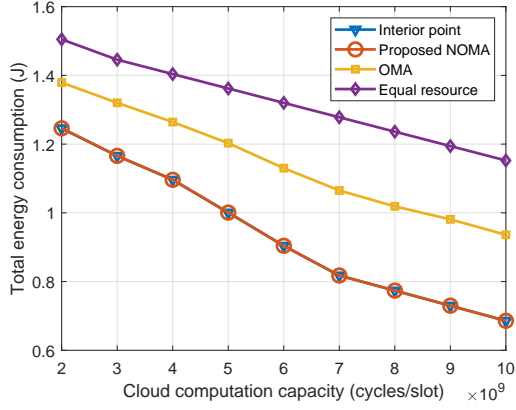}
\vspace{-0.5em}
\caption{Total energy consumption versus cloud computation capacity.}\label{fig3}
\vspace{-0.5em}
\end{figure}

In Fig.~\ref{fig3}, we show the total energy consumption versus cloud computation capacity.
It is observed that the total energy consumption decreases with cloud computation capacity since higher cloud computation capacity allows users to offload more data to the BS, resulting lower energy consumption at users.
The proposed algorithm achieves the best performance according to this figure, which shows the effectiveness of the proposed algorithm.
Besides, the total energy consumption of the proposed NOMA scheme outperforms the conventional OMA scheme, especially when the cloud computation capacity is high.

\section{Conclusion}

In this paper, we have investigated the total energy minimization problem for an uplink NOMA-based MEC network.
The energy minimization problem is shown to be convex.
By analyzing the total energy consumption of all users, we prove that it is optimal to occupy the maximal transmission time.
Besides, we propose an iterative algorithm via solving two subproblems: the time allocation problem and the offloading data allocation problem.
The proposed algorithm is shown to be globally optimal since the time vector and offloading data vector are not coupled in the constraints.
Numerical results show that the proposed algorithm achieves better performance than conventional schemes in terms of energy consumption.


\appendices
\section{Proof of Lemma 1}
\setcounter{equation}{0}
\renewcommand{\theequation}{\thesection.\arabic{equation}}
Since the constraints of problem (\ref{sys1min1}) are all linear, we only need to prove that the objective function (\ref{sys1min1}a) is a convex function.

To show this, we define a function
\begin{equation}\label{AppenAeq1}
f_i (d_{i1}, d_{i2})\!=\!  B \!\left(\!a_{i1} 2^{\frac{d_{i1}+d_{i2}}{B}}
+(a_{i2}-a_{i1}) 2^{\frac{d_{i2}}{B}}
-a_{i2}\!\right)\!,\!
\end{equation}
which is a convex function with respect to (w.r.t.) $(d_{i1}, d_{i2})$ since exponential function is convex, $a_{i1}\geq 0$ and $a_{i2}\geq a_{i1}$ according to (\ref{sys1eq6_1}) and $h_{i1}\geq h_{i2}$.
Based on \cite[Page~89]{boyd2004convex}, the perspective of $u(\pmb x)$ is the function $v(\pmb x,y)$ defined by
\begin{equation}
 v(\pmb x,y)=tu(\pmb x/y), {\textbf{dom}}\:v= \{(\pmb x,y)|\pmb x/y\in {\textbf{dom}}\:u, y>0\}.
 \end{equation}
If $u(\pmb x)$ is a convex function, then so is its perspective function $v(\pmb x,y)$ \cite[Page~89]{boyd2004convex}.
As a result,
\begin{equation}
t_if_i\!\left(\!\frac{d_{i1}}{t_i}, \frac{d_{i2}}{t_i}\!\right)\!=\!B t_i\!\left(\!a_{i1} 2^{\frac{d_{i1}+d_{i2}}{Bt_i}}
+(a_{i2}-a_{i1}) 2^{\frac{d_{i2}}{Bt_i}}
-a_{i2}\!\right)
\end{equation}
is convex w.r.t.
$(d_{i1},d_{i2},t_i)$.
Due to the fact that (\ref{sys1min1}a) is a nonnegative weighted sum of convex functions, the
objective function of problem (\ref{sys1min1}) is convex w.r.t.
$(\pmb d, \pmb t)$ \cite[Page~89]{boyd2004convex}.

\section{Proof of Lemma 2}
\setcounter{equation}{0}
\renewcommand{\theequation}{\thesection.\arabic{equation}}

We first define function
\begin{equation}\label{AppenBeq1}
g_i (t_i)=  B t_i \left(a_{i1} 2^{\frac{d_{i1}+d_{i2}}{Bt_i}}
+(a_{i2}-a_{i1}) 2^{\frac{d_{i2}}{Bt_i}}
-a_{i2}\right).
\end{equation}
Then, we have
\begin{eqnarray}\label{AppenBeq2}
\!\!\!\!\!\!\!\!\!g_i' (t_i)=&&\!\!\!\!\!\!\!\!\!\!\!
  a_{i1}\left(B-{\frac{(\ln 2)(d_{i1}+d_{i2})}{t_i}}\right) 2^{\frac{d_{i1}+d_{i2}}{Bt_i}}
\nonumber\\
&&\!\!\!\!\!\!\!\!\!\!\!
+(a_{i2}-a_{i1})\left(B-{\frac{(\ln 2)d_{i2}}{t_i}}\right) 2^{\frac{d_{i2}}{Bt_i}}
-a_{i2}B.
\end{eqnarray}
From (\ref{AppenBeq2}), we can obtain that
\begin{equation}\label{AppenBeq3}
\lim_{t_i \rightarrow +\infty}g_i'(t_i) =0.
\end{equation}
According to the proof of Lemma 1, $g_i(t_i)$ is a convex function w.r.t. $t_i$, which shows that $g_i''(t_i) > 0$ and $g_i'(t_i)$ is an increasing function.
Combining (\ref{AppenBeq3}) and $g_i'(t_i)$ is an increasing function, we can obtain that $g_i'(t_i)<0$ for all $0<t_i<+\infty$.
As a result, $g_i(t_i)$ is a decreasing function.

We then prove that $\sum_{i=1}^N t_i^*=T$ for the optimal solution to problem (\ref{sys1min1}) by using the contradiction method.
Suppose that the optimal solution $(\pmb d^*, \pmb t^*)$ to problem (\ref{sys1min1}) satisfies $\sum_{i=1}^N t_i^*<T$.
We can increase $t_1^*$ to $\bar t_1=T-\sum_{i=2}^N t_i^*$.
With new solution $(\pmb d^*, \bar {\pmb t}=[\bar t_1, t_2^*,\cdots, t_N^*])$, we can claim that the new solution is feasible with lower objective value, which contradicts that $(\pmb d^*, \pmb t^*)$ is the optimal solution to problem (\ref{sys1min1}).
As a result, Lemma 2 is proved.

\section{Proof of Theorem 1}
\setcounter{equation}{0}
\renewcommand{\theequation}{\thesection.\arabic{equation}}

The Lagrangian function of problem (\ref{sys1min1}) with fixed $\pmb d$ can be written by
\begin{eqnarray}
\!\!\!\!\!\!\!\!\!\mathcal L_1
=&&\!\!\!\!\!\!\!\!\!\!
\sum_{i=1}^N Bt_i\!\left(\!a_{i1} 2^{\frac{d_{i1}+d_{i2}}{Bt_i}}
+(a_{i2}-a_{i1}) 2^{\frac{d_{i2}}{Bt_i}}
-a_{i2}\!\right)
 \nonumber\\
 &&\!\!\!\!\!\!\!\!\!\!+\sum_{i=1}^N \sum_{j=1}^2  (R_{ij}-d_{ij}) C_{ij}P_{ij}+\alpha\left(\sum_{i=1}^N t_i -T\right ),
\end{eqnarray}
where $\alpha$ is a non-negative Lagrangian multiplier associated with constraint (\ref{sys1min1}b).
The first-order derivative of problem (\ref{sys1min1}) with fixed $\pmb d$ can be given by
\begin{equation}\label{AppenCeq1}
\begin{aligned}
\!\!\!\!\!\!\!\!\!\!\!\!&
\frac{\partial \mathcal L_1}{\partial t_{i}}=g_i'(t_i)+\alpha,
\end{aligned}
\end{equation}
where $g_i'(t_i)$ is defined in (\ref{AppenBeq2}).
Setting $\frac{\partial \mathcal L_1}{\partial t_{i}}=0$, we have
\begin{equation}\label{AppenCeq2}
t_i=g_i'^{-1}(-\alpha),
\end{equation}
$g_i'^{-1}(x)$ is the inverse function of the monotonically increasing function $g_i'(x)$.
Considering constraints (\ref{sys1min1}d), the optimal value of $t_i$ is given by (\ref{IA3eq1}).

According to Lemma 2, constraint (\ref{sys1min1}b) holds with equality for the optimal solution.
Substituting (\ref{IA3eq1}) into constraint (\ref{sys1min1}b) with equality yields (\ref{IA3eq2}).
Since $g_i'(x)$ is a monotonically increasing function according to the convexity of objective function (\ref{sys1min1}a), the inverse function $g_i'^{-1}(x)$ is also a monotonically increasing function.
As a result, the unique value of $\alpha$ satisfying (\ref{IA3eq2}) can be  obtained via useing the bisection method.

\section{Proof of Theorem 2}
\setcounter{equation}{0}
\renewcommand{\theequation}{\thesection.\arabic{equation}}

The Lagrangian function of problem (\ref{sys1min1}) with fixed $\pmb t$ can be written by
\begin{eqnarray}
\!\!\!\!\!\!\!\!\!\!\!\!\!\!\!\!\!\!\!\!\!\mathcal L_2 &&\!\!\!\!\!\!\!\!\!\!
=
\sum_{i=1}^N Bt_i\!\left(\!a_{i1} 2^{\frac{d_{i1}+d_{i2}}{Bt_i}}
+(a_{i2}-a_{i1}) 2^{\frac{d_{i2}}{Bt_i}}
-a_{i2}\!\right)
 \nonumber\\
 &&\!\!\!\!\!\!\!\!\!\!\!\!\!\!+\!\sum_{i=1}^N\! \sum_{j=1}^2 \! (R_{ij}\!-\!d_{ij}) C_{ij}P_{ij}\!+\!\beta\!\left(\!\sum_{i=1}^N \!\sum_{j=1}^2\! d_{ij} C_{ij}\! -\! F \! \right)\!\!,\!
\end{eqnarray}
where  $\beta$ is a non-negative Lagrangian multiplier associated with constraint (\ref{sys1min1}c).
The first-order derivatives of problem (\ref{sys1min1}) can be given by
\begin{subequations}\label{AppenDkkt1}
\begin{align}
\!\!\!\!\!\!\!\!\!\!\!\!&
\frac{\partial \mathcal L_2}{\partial d_{i1}}=
(\ln 2)  a_{i1}   2^{\frac{d_{i1}+d_{i2}}{Bt_i}}+(\beta-P_{i1}) C_{i1}
\\
 &\frac{\partial \mathcal L_2}{\partial d_{i2}}=
 (\ln 2)  a_{i1}   2^{\frac{d_{i1}+d_{i2}}{Bt_i}}+ (\ln 2)  (a_{i2}- a_{i1} )  2^{\frac{d_{i2}}{Bt_i}}
 \nonumber\\
 &\qquad\quad+
 (\beta -P_{i2} )C_{i2}.
\end{align}
\end{subequations}

Setting $\frac{\partial \mathcal L_2}{\partial d_{i1}}=0$ into (\ref{AppenDkkt1}a) and $\frac{\partial \mathcal L_2}{\partial d_{i2}}=0$ into (\ref{AppenDkkt1}b), we can obtain
\begin{equation}\label{AppenDeq1}
 d_{i1}=Bt_i\log_2 \!\left(\!
\frac{ (a_{i2}-a_{i1})(  P_{i1}C_{i1}-\beta C_{i1})}
{a_{i1}(\beta(C_{i1}-C_{i2})-P_{i1}C_{i1}+P_{i2}C_{i2} )}
\!\right) \!,
\end{equation}
and
\begin{equation}\label{AppenDeq2}
d_{i2}=Bt_i\log_2 \!\left(\!
\frac{\beta(C_{i1}-C_{i2})-P_{i1}C_{i1}+P_{i2}C_{i2}}
{(\ln 2)(a_{i2}-a_{i1})}
\!\right) \!.\!
\end{equation}
Considering constraints (\ref{sys1min1}d), the value of $d_{i1}$ and $d_{i2}$ are respectively given by (\ref{IA3eq3}) and (\ref{IA3eq5}).

To calculate the value of Lagrange multiplier $\beta$, we consider the following two cases.

1) If $\beta=0$, we can obtain the values of $d_{i1}$ and $d_{i2}$ as in (\ref{IA3eq6}). In this case, constraint (\ref{sys1min1}c) $\sum_{i=1}^N \sum _{j=1}^2 d_{ij} (0) C_{ij} \leq F$ should be satisfied to guarantee the feasibility.

2) If $\beta>0$, constraint (\ref{sys1min1}c) holds with equality according to the complementary slackness condition.
As a result, $\beta$ should satisfy (\ref{IA3eq7}), which can be solved via the one-dimension search method.
For the special case $C_{i1}=C_{i2}$, i.e., the number of CPU cycles required for computing 1-bit input data at users in each group are the same, $d_{i1}(\beta)$ is a monotonically decreasing function w.r.t. $\beta$ and $d_{i2}(\beta)$ is a constant w.r.t. $\beta$.
Thus, the right hand side of equation (\ref{IA3eq7}) is monotonically decreasing, which indicates that (\ref{IA3eq7}) can be effectively solved via the bisection method.

\section{Proof of Theorem 3}
\setcounter{equation}{0}
\renewcommand{\theequation}{\thesection.\arabic{equation}}

We first show that Algorithm 1 converges.
The proof is established by showing that the sum energy value (\ref{sys1min1}a) is non-increasing when the sequence $(\pmb d, \pmb t)$ is updated.
According to the Algorithm 1, we have
\begin{eqnarray}\label{appenEeq1}
V_{\text{obj}}^{(l)}
&&\!\!\!\!\!\!\!\!\!\!\!
=\bar{V}(\pmb d^{(l)},  \pmb t^{(l)})
\nonumber\\&&\!\!\!\!\!\!\!\!\!\!\!
\overset{(\text a)}{\geq}
\bar{V}(\pmb d^{(l)},  \pmb t^{(l+1)})
\nonumber\\&&\!\!\!\!\!\!\!\!\!\!\!
\overset{(\text b)}{\geq}
\bar{V}(\pmb d^{(l+1)},  \pmb t^{(l+1)})=V_{\text{obj}}^{(l+1)},
\end{eqnarray}
where inequality (a) follows from that  $\pmb t^{(l+1)}$ is the optimal time allocation of problem (\ref{sys1min1}) with fixed offloading data ${\pmb d}^{(t)}$,
and inequality (b) follows from that $\pmb d^{(l+1)}$ is the optimal offloading data of problem (\ref{sys1min1}) with fixed time allocation $\pmb t^{(l+1)}$.
Thus, the total energy is non-increasing after the updating of time allocation and offloading data.
Due to that the total energy value (\ref{sys1min1}a) is nondecreasing in each step from (\ref{appenEeq1}) and the total energy value (\ref{sys1min1}a) is finitely lower-bounded (positive), Algorithm 1 must converge.

We then show that the convergent solution of Algorithm 1 is the globally optimal solution to problem (\ref{sys1min1}).
The Lagrangian function of problem (\ref{sys1min1}) can be written by
\begin{eqnarray}
\!\!\!\!\!\!\!\!\!\mathcal L
=&&\!\!\!\!\!\!\!\!\!\!
\sum_{i=1}^N Bt_i\!\left(\!a_{i1} 2^{\frac{d_{i1}+d_{i2}}{Bt_i}}
+(a_{i2}-a_{i1}) 2^{\frac{d_{i2}}{Bt_i}}
-a_{i2}\!\right)
 \nonumber\\
 &&\!\!\!\!\!\!\!\!\!\!+\sum_{i=1}^N \sum_{j=1}^2  (R_{ij}-d_{ij}) C_{ij}P_{ij}+\alpha\left(\sum_{i=1}^N t_i -T\right )
  \nonumber\\
 &&\!\!\!\!\!\!\!\!\!\!+\beta\left(\sum_{i=1}^N \sum_{j=1}^2 d_{ij} C_{ij} - F  \right),
\end{eqnarray}
where $\alpha$ and $\beta$ are non-negative Lagrangian multipliers associated with constraints (\ref{sys1min1}b) and (\ref{sys1min1}c), respectively.

Denote $(\pmb d^*, \pmb t ^*)$ as the solution obtained by Algorithm 1.
There exist $\alpha$ and $\beta$ such that
\begin{equation}\label{appenEeq3}
\left.\frac{\partial \mathcal L}{\partial d_{ij}}\right|_{d_{ij}=d_{ij}^*}\left\{
\begin{array}{ll}
\!\!=0& \text{if} \;d_{ij}^*\in(D_{ij},R_{ij})\\
\!\!\geq 0& \text{if} \;d_{ij}^*=D_{ij} \\
\!\!\leq 0& \text{if} \; d_{ij}^* = R_{ij}
\end{array}
\right.
\end{equation}
for all $i \in \mathcal N, j=1,2$,
and
\begin{equation}\label{appenEeq6}
\left.\frac{\mathcal L}{\partial t_i}\right|_{t_i=t_i^*}\left\{
\begin{array}{ll}
\!\!=0& \text{if} \;t_i^*>0\\
\!\!\geq 0& \text{if} \;t_i^*=0
\end{array}
\right.,\quad\forall i \in \mathcal N,
\end{equation}
since $\pmb d^*$ is the optimal solution of problem (\ref{sys1min1}) with fixed $\pmb t^*$ and $\pmb t^*$ is the optimal solution of problem (\ref{sys1min1}) with given $\pmb d^*$.
According to (\ref{appenEeq3}) and (\ref{appenEeq6}), solution $(\pmb d^*, \pmb t^*)$ satisfies the KKT conditions of problem (\ref{sys1min1}), i.e., the locally optimal solution $(\pmb d^*, \pmb t^*)$ is the globally optimal solution to convex problem (\ref{sys1min1}).

\section*{Acknowledgment}
This work was supported by the Engineering and Physical Science Research Council (EPSRC) through the Scalable Full Duplex Dense Wireless Networks (SENSE) grant EP/P003486/1.

\bibliographystyle{IEEEtran}
\bibliography{IEEEabrv,MMM}
\end{document}